%% file: main.tex
\documentclass[sigplan,10pt,nonacm]{acmart}

\AtBeginDocument{%
  }

\usepackage{algorithm}
\usepackage{xspace}
\usepackage{algpseudocode}
\usepackage{subcaption}
\usepackage{fontawesome}
\usepackage{listings}
\usepackage[cachedir=minted_cache]{minted}
\usepackage{pifont}
\usepackage{tikz}
\usepackage[most]{tcolorbox}

\newcommand{\circnumblack}[1]{%
\tikz[baseline=(char.base)]{
\node[shape=circle, fill=black, text=white, inner sep=1pt] (char) {#1};
}}

\begin{document}

\title{Torcalia.jl: An MPI-based Runtime System for Asynchronous Task Parallelism in Julia}

\author{Argiris Sofotasios}
\email{a.sofotasios@ceid.upatras.gr}
\orcid{0009-0002-5282-8064}
\affiliation{%
  \institution{Department of Computer Engineering and Informatics, University of Patras}
  \city{Patras}
  \country{Greece}
}
\affiliation{%
  \institution{Archimedes Unit, Athena Research Center}
  \city{Athens}
  \country{Greece}
}

\author{Dimitris Metaxakis}
\email{d.metaxakis@ceid.upatras.gr}
\orcid{0009-0005-4452-891X}
\affiliation{%
  \institution{Department of Computer Engineering and Informatics, University of Patras}
  \city{Patras}
  \country{Greece}
}
\affiliation{%
  \institution{Archimedes Unit, Athena Research Center}
  \city{Athens}
  \country{Greece}
}

\author{Chrysostomos Stylios}
\email{stylios@athenarc.gr}
\orcid{0000-0002-2888-6515}
\affiliation{%
  \institution{Department of Informatics and Telecommunications, University of Ioannina}
  \city{Arta}
  \country{Greece}
}
\affiliation{%
  \institution{Industrial Systems Institute, Athena Research Center}
  \city{Patras}
  \country{Greece}
}

\author{Panagiotis Hadjidoukas}
\email{phadjido@ceid.upatras.gr}
\orcid{0000-0002-2528-7568}
\affiliation{%
  \institution{Department of Computer Engineering and Informatics, University of Patras}
  \city{Patras}
  \country{Greece}
}
\affiliation{%
  \institution{Industrial Systems Institute, Athena Research Center}
  \city{Patras}
  \country{Greece}
}


\begin{abstract}
Designed to address the two-language problem, the Julia programming language provides a powerful platform for developing high-performance scientific computing and machine learning applications. Such applications increasingly exhibit irregular, adaptive, or data-driven workloads for which traditional bulk-synchronous parallel (BSP) models often struggle to efficiently exploit the available parallelism. Asynchronous many-task (AMT) runtime systems offer a compelling alternative, decomposing computations into fine-grained tasks that are dynamically scheduled across the hierarchical architecture of modern HPC systems. 

This paper introduces Torcalia.jl, a Julia package designed to orchestrate the asynchronous execution of tasks across both shared and distributed memory environments. Built on top of MPI.jl and Julia's native multithreading, Torcalia employs a dependency-aware task scheduling scheme that combines the work-sharing and work-stealing paradigms to achieve adaptive load balancing. Tasks can be nested arbitrarily, enabling applications to exploit multiple levels of parallelism, while application-level distributed shared memory (DSM) extends task-based shared-memory programming to multi-node clusters. Beyond task-based programming, the proposed library supports skeletal programming, providing high-level abstractions that facilitate the implementation of parallel applications at scale. Evaluated on the ARIS supercomputer using four diverse workloads, Torcalia demonstrates strong scalability, reducing wall-clock time by up to 18.1\% compared with Dagger, a state-of-the-art task-based engine in Julia. The source code of Torcalia.jl is publicly available at~\href{https://github.com/CEID-HPCLAB/Torcalia.jl}{\faGithub}.

\end{abstract}

\maketitle

\input{./sections/motivation.tex}
\input{./sections/software.tex}
\input{./sections/examples.tex}
\input{./sections/evaluation.tex}
\input{./sections/impact.tex}
\input{./sections/conclusion.tex}
\input{./sections/acks.tex}
\input{./sections/funding.tex}

\bibliographystyle{ACM-Reference-Format}
\bibliography{references}

\end{document}

%% file: sections/motivation.tex
\section{Motivation and Significance}
\label{sec:motivation}

With Moore's Law significantly slowing down, the ever-increasing demand for computational power has made cluster environments, consisting of multi- and many-core processing units, the prevalent platform for high-performance computing (HPC). However, the growing heterogeneity of modern parallel computing systems, in both hardware and software~\cite{nesi2023, agullo2017}, poses significant programmability and portability challenges~\cite{Reitz2025}, often hindering efficient resource utilization. In this evolving landscape, task-based programming models have emerged as a promising approach to address these challenges, providing a high-level paradigm for expressing parallelism across the hierarchical architecture of contemporary HPC systems~\cite{Reitz2025, agullo2017, alomairy2020, Posner2025}.

In task-based programming, applications are expressed as collections of asynchronous fine-grained tasks that can be spawned at different levels of parallelism and executed across the processing units of a single node, a cluster, or a supercomputer~\cite{torcpy, Reitz2025, Posner2025}. The tasks can be completely decoupled or part of an iterative algorithm in which the results of each step guide the next. Their execution is delegated to a runtime system, which dynamically schedules individual tasks across the available resources while respecting the underlying dependency graph~\cite{Taxonomy}. This allows programmers to focus on the problem at hand, abstracting away low-level execution concerns such as load balancing, synchronization, data movement, and communication~\cite{Dastur2026, Joseph2026}. Such an abstraction is particularly attractive for irregular, adaptive, or non-iterative workloads, where parallelism is difficult to express through static loop-based or message-passing approaches~\cite{bantel2026, Reitz2025, Gao2025, HRPF, Posner2025}. As such, task-based models have been widely adopted to accelerate applications in various fields~\cite{agullo2023, Nesi2020, valois2026, asgari2023}. 

Designed to combine the performance of C, the ease of Python, and the statistical capabilities of R, Julia is a dynamic language for high-performance scientific computing~\cite{Julia}. Built on the LLVM compiler infrastructure, Julia uses just-in-time (JIT) compilation together with type inference, code specialization, and multiple dispatch to achieve performance comparable to that of statically typed languages~\cite{Giordano2022, Dagger}. These features, combined with its high-level syntax, extensive standard library, and rich package ecosystem, provide a unified environment that bridges the gap between productivity and performance, addressing the \textit{two-language problem}~\cite{Julia2018}. As a result, Julia has emerged as a powerful platform for developing scientific computing and machine learning frameworks~\cite{jubat, Dunning2017, Luo2022, Flux}, which often involve complex workflows, where traditional bulk-synchronous parallel (BSP) paradigms struggle to efficiently express and exploit available parallelism~\cite{Joseph2026}. Addressing these challenges has led to the development of asynchronous many-task (AMT) runtime systems in Julia~\cite{Dagger, Dastur2026}. 

However, existing task-based engines in Julia are primarily designed for general-purpose parallel programming, which can pose performance and scalability challenges when applied to tightly coupled HPC workloads characterized by irregular task granularity, nested parallelism, and highly dynamic execution behavior. Furthermore, they lack direct interoperability with applications based on the Message Passing Interface (MPI), the \textit{lingua franca} for inter-node communication in HPC, limiting their adoption in large-scale applications. Dynamic switching to the native Single Program Multiple Data (SPMD) execution model of MPI is also not supported. Another limitation is that most of them require users to specify the available workers \emph{a priori}, without supporting their dynamic adjustment at runtime. In addition, they provide limited fine-grained control over task scheduling and load balancing, which prevents expert users from incorporating application-specific knowledge into runtime decisions. Finally, their APIs lack support for parallel \emph{algorithmic skeletons}~\cite{cole2004}, increasing the programming effort required to express common data-parallel computation patterns efficiently.

To overcome these limitations, we introduce \textbf{Torcalia.jl}, a platform-independent package for AMT parallelism in Julia. Built on top of MPI and Julia's native multithreading, Torcalia employs a completely decentralized model to dynamically orchestrate the execution of tasks on both shared and distributed memory platforms. To ensure efficient resource utilization, adaptive load balancing is achieved through a hybrid scheduling scheme that combines the \emph{work-sharing}~\cite{Taxonomy} and \emph{work-stealing}~\cite{workstealing} paradigms, while also supporting task pinning and prioritization. At the application level, the proposed tasking library exposes a high-level API inspired by the Python Enhancement Proposal (PEP) 3148~\cite{pep3148}, abstracting task creation and synchronization through simple \texttt{submit} and \texttt{wait} calls. Moreover, a comprehensive set of parallel algorithmic skeletons is provided, enabling researchers from diverse domains to seamlessly scale their workloads from local workstations to multi-node 
clusters. 

\vspace{0.2em}

\noindent Overall, Torcalia aims to extend Julia's HPC ecosystem by providing a portable AMT runtime system that:

\vspace{-0.3em}

\begin{itemize}
    
    \item Offers an intuitive programming model that simplifies the expression of asynchronous task-based parallelism across shared and distributed memory environments.

    \item Adopts an MPI+X model, transparently leveraging MPI primitives for inter-node communication and Julia-native multithreading for intra-node parallelism, while allowing MPI operations to be invoked at the application level. Dynamic switching to the standard SPMD execution model is also supported. 

    \item Supports nested parallelism, facilitating the implementation of parallel recursive algorithms.

    \item Automatically detects and orchestrates data dependencies between tasks. 

    \item Integrates inter-node work stealing for load balancing, enhanced by a set of task distribution policies that can be customized by the user.

    \item Provides high-level functions that enable the use of software-distributed shared memory (S-DSM) at the application level.

    \item Enables \emph{on-demand} spawning and termination of MPI worker processes at runtime.

    \item Exposes the above features through a user-friendly API that encapsulates asynchronous task execution via \emph{futures} (task descriptors).

\end{itemize}

\vspace{-0.3em}

\begin{figure*}[t!]
\centering
\includegraphics[width=0.9\textwidth]{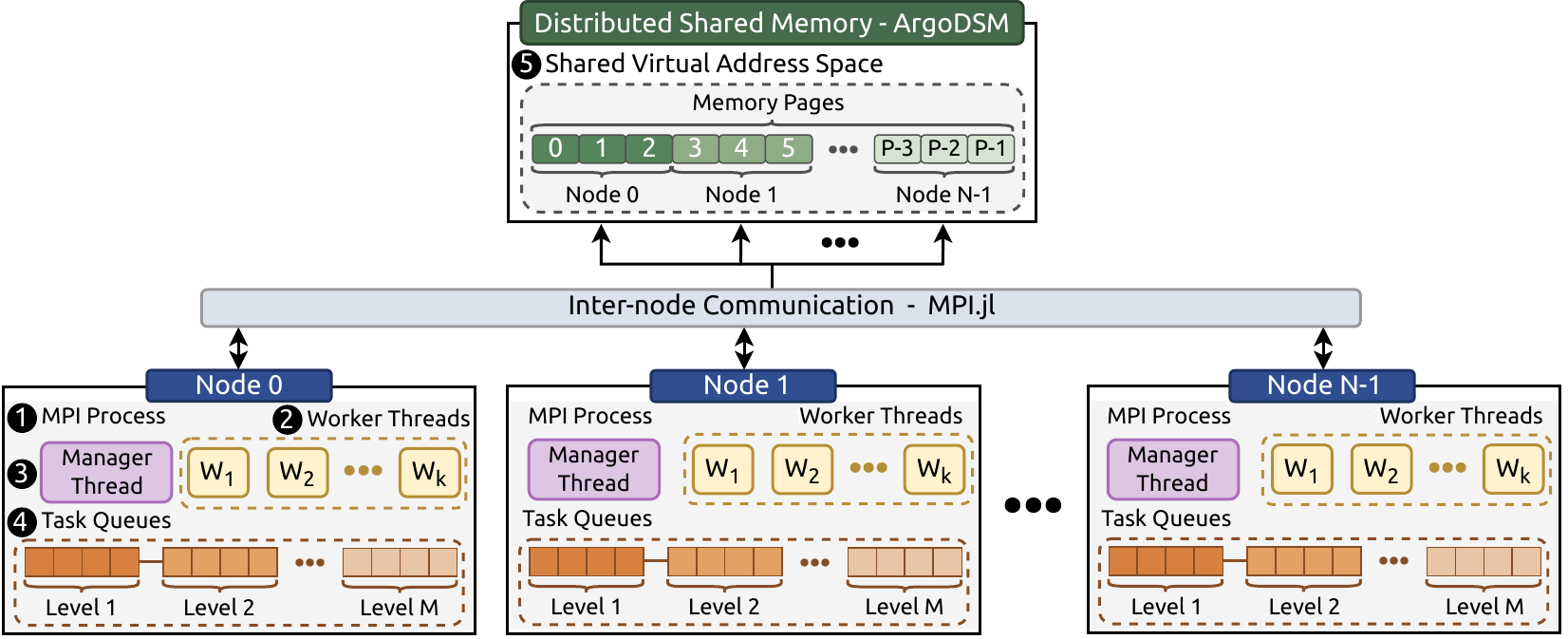}
\caption{High-level architecture of the Torcalia AMT runtime system.}
\vspace{-0.2cm}
\label{fig:architecture}
\end{figure*}

The core execution model of Torcalia is inspired by TORC \cite{torc}, a C runtime library for task-based parallelism on heterogeneous clusters. TORC has been employed for a wide range of applications, including optimization~\cite{voglis2013, verma2017}, uncertainty quantification~\cite{pi4u, phadjido2018}, and numerical differentiation~\cite{ndl}. A high-level Python implementation, namely torcpy~\cite{torcpy}, has also been developed and primarily used in machine learning workflows~\cite{DataLoader, KIPO}. The proposed tasking library combines the core principles of TORC with the programmability of torcpy while introducing a set of new features, including parallel algorithmic skeletons, task prioritization, and software-based DSM support, providing a robust AMT runtime system for Julia that can be adopted across a broad range of large-scale workloads.

The remainder of this paper is organized as follows. Subsection~\ref{subsec:related-work} overviews Julia’s parallel computing ecosystem and presents existing efforts towards AMT runtime systems. Section~\ref{sec:software} introduces Torcalia, providing a detailed description of its runtime engine, programming model, and supported features, while Section~\ref{sec:examples} demonstrates the use of the package in four illustrative use cases. Section~\ref{sec:evaluation} evaluates the scalability of Torcalia on a multi-node cluster and compares its performance with Dagger~\cite{Dagger}, the state-of-the-art framework for task-based parallelism in Julia. Section~\ref{sec:impact} discusses the impact of Torcalia on Julia's HPC ecosystem, and Section~\ref{sec:conclusion} concludes the article and outlines future directions.

\vspace{-0.3em}

\subsection{Related Work}
\label{subsec:related-work}

Originally tailored for high-performance scientific computing, Julia provides built-in support for a variety of parallel computing paradigms. For shared-memory architectures, multithreading is supported through the \texttt{Threads} module. The \texttt{@spawn} and \texttt{@threads} macros enable asynchronous task creation and parallel loop execution, respectively, while the \texttt{wait} and \texttt{fetch} routines provide mechanisms for task synchronization and result retrieval. For distributed-memory computing, Julia provides the \texttt{Distributed} module that enables parallel computing across multiple processes, either on a single machine or across a multi-node cluster~\cite{Dagger, Dastur2026}. Its programming model is built around two primitives, \emph{remote references} and \emph{remote calls}, with communication implemented over standard TCP. Although portable, this approach cannot leverage specialized low-latency interconnects such as InfiniBand, thereby limiting attainable performance in HPC clusters~\cite{MPI.jl, guimaraes2025}.  

Each of the aforementioned paradigms follows a different parallelism approach and employs its own API to exploit parallelism~\cite{Dagger}. As a result, combining their strengths within a hybrid execution model introduces challenges, particularly for users without HPC expertise. To address this, Alomairy \textit{et al.} introduced Dagger~\cite{Dagger}, a unified framework for dynamic task scheduling in heterogeneous environments. Dagger builds on the \texttt{Distributed} module and adopts a sequential, macro-based programming model, enabling users to spawn tasks using straightforward Julia code. Drawing inspiration from Dask~\cite{Dask}, it relies on a directed acyclic graph (DAG) scheduler that automatically detects and orchestrates data dependencies between tasks, while managing their distribution across available nodes. Our experiments indicate that Dagger's scheduler can struggle to distribute work efficiently, particularly in the presence of irregular workloads~(§\ref{sec:evaluation}). 

To enable scalable execution on HPC clusters, Guimarães \textit{et al.} developed an MPI-based backend for Dagger~\cite{guimaraes2025}. Their experimental evaluation showed that, at large node counts, the proposed backend exhibits limited scalability and is consistently outperformed by the original implementation. As the authors acknowledge, the communication layer requires further optimization for large-scale HPC environments. More recently, Dastur \textit{et al.} proposed DistributedWorkflows~\cite{Dastur2026}, a high-level wrapper around GPI-Space~\cite{GPI-Space} that enables researchers to express application parallelism through Petri net workflows. Although it offers an expressive programming model, DistributedWorkflows has several limitations. As an interface package, it relies on external tools beyond Julia and entails additional setup complexity. Moreover, it requires a shared file system whose high I/O overhead limits its applicability to long-running tasks~\cite{Dastur2026}.

%% file: sections/software.tex
\vspace{-0.3em}

\section{Software Description}
\label{sec:software}

\textbf{Torcalia} provides a Julia-native platform for asynchronous task-based parallelism on both shared and distributed memory environments. As shown in Figure~\ref{fig:architecture}, its runtime engine is based on a decentralized model, on top of which a hybrid MPI+X parallelism scheme is employed for the concurrent execution of tasks. Inter-node communication relies on the \texttt{MPI.jl} package~\cite{MPI.jl}, which provides Julia bindings for MPI, while intra-node parallelism is supported by Julia's \texttt{Threads} module. Within each MPI process, inter-thread coordination is achieved through thread-safe queues provided by the \texttt{ConcurrentCollections.jl} package~\cite{ConcurrentCollections}. 

In Torcalia, an application is abstracted into a hierarchy of tasks, with user code serving as the top-level task. Parallelism is expressed through task submissions, each defined by a user-provided function with optional positional and/or keyword arguments. Arguments may include \emph{futures} referring to the results of previously submitted tasks, thereby expressing data dependencies. The runtime system automatically detects these dependencies and dynamically schedules ready tasks across available processing units to exploit concurrency, ensuring load balancing. Explicit synchronization between tasks is provided through high-level functions, enabling users to precisely coordinate task execution when required. Execution parameters, such as processor affinity and priority level, can be specified in task submissions, providing fine-grained control over task scheduling.

\vspace{0.598em}

\noindent\textbf{Execution Model.} A Torcalia-based application consists of multiple MPI processes~\circnumblack{1}, each hosting a fixed-size pool of Julia worker threads~\circnumblack{2}. In addition to these threads, each MPI process maintains a dedicated \emph{manager thread}~\circnumblack{3} that handles all remote task-coordination operations, such as serving incoming task stealing requests. These operations are performed asynchronously using explicit MPI primitives that remain completely transparent to the user. Note that all MPI processes participate in task execution without a dedicated master process. Using MPI-2 dynamic process management routines~\cite{MPI-2}, the number of MPI processes can be dynamically adjusted at runtime, which is particularly useful for workloads with time-varying computational demands. Moreover, tasks can be spawned at multiple parallelism levels from any participating MPI process, enabling nested parallelism. In addition to task-based execution, Torcalia supports MPI's SPMD programming model through the \texttt{spmd} function, allowing both paradigms to coexist within the same application.

\vspace{0.6em}

\noindent\textbf{Task Scheduling.} Upon task submission, a scheduling policy determines the MPI rank to which each task is assigned. Multiple policies are provided, including round-robin and weighted round-robin. User-defined policies are also supported, allowing application-specific knowledge to be incorporated into runtime decisions. Each MPI rank maintains hierarchical task queues~\circnumblack{4}, into which the assigned ready-to-run tasks are placed according to their priority and spawning level. Worker threads continuously extract tasks from these local queues, with high-priority tasks served first, process each task upon retrieval, and return the corresponding result to the manager thread of the MPI rank that spawned the task. Among \emph{general-purpose} tasks, i.e., those assigned the default priority, tasks from higher spawning levels (\textit{coarse-grained} tasks) are prioritized, while tasks at the same level are retrieved in FIFO order. By decoupling workers from tasks, worker threads never block within the dispatch loop. Whenever a task reaches a synchronization point and waits for the completion of previously submitted tasks, the worker suspends its execution and proceeds with another pending task. Once the awaited tasks have been completed, the worker resumes execution of the suspended task.

\noindent\textbf{Work Stealing.} Beyond \emph{work sharing}, Torcalia also employs a \emph{work-stealing} mechanism that follows an \emph{on-demand} approach, initiating work stealing only when needed, without requiring global knowledge of system load. Specifically, an idle worker thread first attempts to retrieve work from its local workpile. If no tasks are available locally, it issues a task-stealing request to the manager thread of the next MPI rank. If the request yields no work, the worker proceeds to the next rank and repeats the process until a pending task is obtained. This \textit{work-first, help-then} policy provides adaptive load balancing, reducing processor idle time even for highly irregular workloads. 

\vspace{-0.3em}

\subsection{Software-Distributed Shared Memory (S-DSM)}
\label{sec:sdsm}

DSM systems provide coherent global address spaces across distributed memory environments, extending the shared-memory programming model beyond the scale of single-node systems. Page-based DSM (PDSM) is a common approach to S-DSM that implements the shared memory abstraction through the virtual memory system. Torcalia builds on ArgoDSM~\cite{argoDSM}, a state-of-the-art PDSM system, to provide application-level support for S-DSM~\circnumblack{5}. Users need only explicitly allocate the corresponding shared memory regions through the \texttt{pdsm\_alloc} function. The runtime transparently manages the underlying communication and synchronization using the primitives provided by ArgoDSM, allowing the allocated data to be accessed within tasks as in shared-memory programming. Beyond PDSM, Torcalia also supports the \emph{MPI shared memory} paradigm, enabling processes within a shared memory island to directly map memory regions into a global shared address space. Shared-memory windows are allocated through the \texttt{shw\_alloc} function, while the underlying MPI-3 RMA operations~\cite{mpi-3} are handled internally by the runtime. This paradigm can reduce memory consumption by avoiding data replication across tasks running on different MPI processes.

\vspace{-0.3em}

\subsection{Parallel Algorithmic Skeletons}
\label{sec:skeletons}

Algorithmic skeletons are generic, high-level programming constructs that abstract common data-parallel computation patterns~\cite{cole2004}. Derived from higher-order functions, they can be parameterized with application-specific sequential code while their parallel orchestration is encapsulated by a runtime system. By composing skeletons, users can build structured, high-level specifications of complex parallel workflows. Torcalia provides built-in support for seven such skeletons, including \emph{map}, \emph{filter}, and \emph{reduce}. Each skeleton is exposed as a function that typically takes two main arguments: one or more iterable objects containing the input data and a user-defined function specifying the operation to be applied to them. Internally, the runtime system decomposes the underlying computation pattern into a set of asynchronous tasks and transparently orchestrates their parallel execution. Skeleton composition is supported either sequentially by passing intermediate results between skeletons or hierarchically by nesting one skeleton within another. To the best of our knowledge, Torcalia is the first Julia-native AMT runtime system to support \emph{skeletal programming} across both shared and distributed memory environments.

\vspace{-0.3em}

\subsection{Application Programming Interface (API)}
\label{sec:api}

Inspired by PEP 3148, Torcalia exposes a high-level interface similar to Python's \texttt{concurrent.futures} module. Tasks are submitted through the \texttt{submit} function, which asynchronously schedules their execution and returns a \emph{future} representing the pending result of the task function. Execution parameters, including the scheduling policy for the submitted task, can be specified as keyword arguments to \texttt{submit}. Upon task completion, the result is stored in the corresponding \emph{future} object and can be retrieved using the \texttt{result} method. Synchronization is achieved through the \texttt{wait} function, which blocks the execution of the calling task until a specified set of previously submitted tasks has been completed. Skeletons are exposed through dedicated functions, while the number of MPI processes can be dynamically adjusted using the \texttt{add\_worker} and \texttt{remove\_worker} functions. Additional routines are provided for engine setup, querying the runtime environment, configuring the scheduling policy, and controlling the work-stealing mechanism.  

%% file: sections/examples.tex
\section{Illustrative Examples}
\label{sec:examples}

\definecolor{backcolour}{rgb}{0.96,0.96,0.96}
\definecolor{blue}{rgb}{0.20,0.32,0.68}
\definecolor{orange}{HTML}{C15F00}
\definecolor{magenta}{HTML}{920035}
\definecolor{navy}{HTML}{243B6B}
\definecolor{teal}{HTML}{155E63}
\definecolor{grey}{HTML}{626872}
\definecolor{purple}{HTML}{542A7A}
\definecolor{kwgreen}{HTML}{008000}

\begin{figure}[!b]
\vspace{-1.2em}
\centering
    \begin{tcolorbox}[
        colback = backcolour,
        colframe = black!10,
        arc = 1.5mm,
        boxrule = 0.25pt,
        left = 0.5mm,
        right = 0.5mm,
        top = 0.28mm,
        bottom = 0.28mm,
        listing only,
    ]
\begin{minted}{julia}
import Torcalia |\textcolor{kwgreen}{\textbf{as}}| Torc

const Future = Torc.|{\color{magenta}Future}|

|{\color{navy}square}|(|{\color{orange}x}|::|{\textbf{\color{blue}Int}}|)::|{\textbf{\color{blue}Int}}| = |{\color{orange}x}|^2
|{\color{navy}double}|(|{\color{orange}x}|::|{\textbf{\color{blue}Future}}|)::|{\textbf{\color{blue}Int}}| = 2 * Torc.|{\color{magenta}result}|(|{\color{orange}x}|)

function |{\color{navy}add}|(|{\color{orange}x}|::|{\textbf{\color{blue}Future}}|, |{\color{orange}y}|::|{\textbf{\color{blue}Future}}|)::|{\textbf{\color{blue}Int}}|
    Torc.|{\color{magenta}result}|(|{\color{orange}x}|) + Torc.|{\color{magenta}result}|(|{\color{orange}y}|)
end

function |{\color{navy}main}|()
    taskA = Torc.|{\color{magenta}submit}|(|{\color{navy}square}|, 2)
    taskB = Torc.|{\color{magenta}submit}|(|{\color{navy}square}|, 4)

    taskC = Torc.|{\color{magenta}submit}|(|{\color{navy}add}|, taskA, taskB)
    taskD = Torc.|{\color{magenta}submit}|(|{\color{navy}double}|, taskA)

    Torc.|{\color{magenta}wait}|([taskC, taskD])
    
    @assert Torc.|{\color{magenta}result}|(taskC) == 20
    @assert Torc.|{\color{magenta}result}|(taskD) == 8
end

Torc.|{\color{magenta}init}|(|{\color{navy}main}|)
\end{minted}
    \end{tcolorbox}
\vspace{-1.02em}
\captionof{listing}{Dependency-aware dynamic task scheduling.}
\label{lst:deps}
\vspace{-0.3em}
\end{figure}

The following four code examples illustrate representative use cases of Torcalia, each demonstrating a distinct feature of its runtime engine. For a complete description of the API and all supported features (e.g., task prioritization, switching to the SPMD execution model), we refer the reader to ~\href{https://github.com/CEID-HPCLAB/Torcalia.jl}{\faGithub}.

\vspace{0.6em}

\noindent\textbf{Task Dependencies.} Listing~\ref{lst:deps} illustrates how data dependencies between tasks are expressed in Torcalia, enabling dependency-aware dynamic task scheduling. Task $C$ depends on the results of Tasks $A$ and $B$ and is therefore scheduled only after both complete, whereas Task $D$ depends only on Task $A$ and can start as soon as $A$ completes. Tasks $A$ and $B$ can be executed concurrently, as can Tasks $C$ and $D$, since they are independent of each other. Note that dependencies can also be expressed implicitly by passing data structures containing tasks (e.g., arrays) as arguments.

\vspace{0.6em}

\noindent\textbf{Nested Parallelism.} Each term of the Fibonacci sequence is defined as the sum of the two preceding terms and can thus be computed recursively. Listing~\ref{lst:fib} demonstrates the parallel recursive computation of the $40$th term using Torcalia. Multiple levels of parallelism are exposed by recursively submitting two tasks within each task function and waiting for their results, forming a hierarchical dependency graph. To avoid task oversubscription, the depth of recursive task spawning is bounded. Since this example is intended to show Torcalia's support for nested parallelism, identical function calls are recomputed rather than memoized.

\begin{figure}[!t]
\centering
    \begin{tcolorbox}[
        colback = backcolour,
        colframe = black!10,
        arc = 1.5mm,
        boxrule = 0.25pt,
        left = 0.5mm,
        right = 0.5mm,
        top = 0.28mm,
        bottom = 0.28mm,
        listing only,
    ]
\begin{minted}{julia}
import Torcalia |\textcolor{kwgreen}{\textbf{as}}| Torc

function |{\color{navy}fibonacci}|(|{\color{orange}n}|::|{\textbf{\color{blue}Int}}|)::|{\textbf{\color{blue}Int}}|
    |{\color{orange}n}| < 2 && return |{\color{orange}n}|
    |{\color{orange}n}| < 30 && return |{\color{navy}fibonacci}|(|{\color{orange}n}| - 1) + |{\color{navy}fibonacci}|(|{\color{orange}n}| - 2)

    taskA = Torc.|{\color{magenta}submit}|(|{\color{navy}fibonacci}|, |{\color{orange}n}| - 1)
    taskB = Torc.|{\color{magenta}submit}|(|{\color{navy}fibonacci}|, |{\color{orange}n}| - 2)

    Torc.|{\color{magenta}wait}|([taskA, taskB])
    Torc.|{\color{magenta}result}|(taskA) + Torc.|{\color{magenta}result}|(taskB)
end

function |{\color{navy}main}|()
    N = 40; value = |{\color{navy}fibonacci}|(N)

    @assert value == 102_334_155
end

Torc.|{\color{magenta}init}|(|{\color{navy}main}|)
\end{minted}
    \end{tcolorbox}
\vspace{-1.02em}
\captionof{listing}{Parallel recursive computation of the $40$th Fibonacci number.}
\label{lst:fib}
\vspace{-1em}
\end{figure}

\vspace{0.6em}

\noindent\textbf{Software-Distributed Shared Memory.} A short example of how the DSM abstraction can be used in Torcalia is shown in Listing~\ref{lst:pdsm}. The primary application task allocates and initializes the shared memory region $A$, which is accessible to tasks running on different MPI nodes. Then, $N$ tasks are created, where $N$ is the number of MPI nodes used to launch Torcalia. Each task is explicitly scheduled to a different MPI node and executes the \texttt{work} function, which writes the MPI rank of its assigned node to the corresponding position in $A$.

\begin{figure}[!h]
\centering
    \begin{tcolorbox}[
        colback = backcolour,
        colframe = black!10,
        arc = 1.5mm,
        boxrule = 0.25pt,
        left = 0.5mm,
        right = 0.5mm,
        top = 0.28mm,
        bottom = 0.28mm,
        listing only,
    ]
\begin{minted}{julia}
import Torcalia |\textcolor{kwgreen}{\textbf{as}}| Torc

function |{\color{navy}work}|(|{\color{orange}A}|::|{\textbf{\color{blue}Vector}}|{|{\textbf{\color{blue}Int}}|})
    rank = Torc.|{\color{magenta}node\_id}|(); |{\color{orange}A}|[rank + 1] = rank
    Torc.|{\color{magenta}argo\_barrier}|(1)
end

function |{\color{navy}main}|()
    N = Torc.|{\color{magenta}num\_nodes}|()
    A = Torc.|{\color{magenta}pdsm\_alloc}|(N, |{\textbf{\color{blue}Int}}|)

    for i in 1:N A[i] = 0 end

    tasks = |{\textbf{\color{blue}Vector}}|{|{\textbf{\color{blue}Torc.Future}}|}(undef, N)    
    for i in 0:N-1
        tasks[i+1] = Torc.|{\color{magenta}submit}|(|{\color{navy}work}|, A; |{\color{teal}node}| = i)
    end

    Torc.|{\color{magenta}wait}|(tasks)

    @assert A == |{\color{purple}collect}|(|{\textbf{\color{blue}Int}}|, 0:N-1)
end

|{\color{grey}\char35\ Configure the ArgoDSM memory and cache capacities}|
Torc.|{\color{magenta}argo\_init}|(1024*1024, 1024*4)
Torc.|{\color{magenta}init}|(|{\color{navy}main}|)
\end{minted}
    \end{tcolorbox}
\vspace{-1.02em}
\captionof{listing}{Integrating ArgoDSM into Torcalia for S-DSM across tasks on distinct MPI nodes.}
\label{lst:pdsm}
\vspace{-1em}
\end{figure}

\noindent\textbf{Algorithmic Skeletons.} The proposed tasking library provides seven parallel algorithmic skeletons, four of which are included in Listing~\ref{lst:skeletons}. In this example, all four skeletons operate on the same input vector $A$: \emph{map} computes element-wise squares, \emph{reduce} sums the elements, \emph{mapReduce} combines both operations to compute the sum of the squared elements, and \emph{filter} selects the odd elements. By default, skeleton invocations include an implicit barrier that waits for all spawned tasks to complete. Consequently, in this example, no explicit \texttt{wait} call is required.

\begin{figure}[!t]
\centering
    \begin{tcolorbox}[
        colback = backcolour,
        colframe = black!10,
        arc = 1.5mm,
        boxrule = 0.25pt,
        left = 0.5mm,
        right = 0.5mm,
        top = 0.28mm,
        bottom = 0.28mm,
        listing only,
    ]
\begin{minted}{julia}
import Torcalia |\textcolor{kwgreen}{\textbf{as}}| Torc

|{\color{navy}square}|(|{\color{orange}x}|::|{\textbf{\color{blue}Int}}|)::|{\textbf{\color{blue}Int}}| = |{\color{orange}x}|^2
|{\color{navy}add}|(|{\color{orange}x}|::|{\textbf{\color{blue}Int}}|, |{\color{orange}y}|::|{\textbf{\color{blue}Int}}|)::|{\textbf{\color{blue}Int}}| = |{\color{orange}x}| + |{\color{orange}y}|
|{\color{navy}is\_odd}|(|{\color{orange}x}|::|{\textbf{\color{blue}Int}}|)::|{\textbf{\color{blue}Bool}}| = |{\color{orange}x}| % 2 == 1

function |{\color{navy}main}|()
    N = 8 
    A = |{\color{purple}collect}|(1:N) |{\color{grey}\char35\ [1, 2, 3, 4, 5, 6, 7, 8]}|

    squares = Torc.|{\color{magenta}map}|(|{\color{navy}square}|, A)
    total_sum = Torc.|{\color{magenta}reduce}|(|{\color{navy}add}|, A)
    sum_squares = Torc.|{\color{magenta}mapReduce}|(|{\color{navy}square}|, |{\color{navy}add}|, A) 
    odds = Torc.|{\color{magenta}filter}|(|{\color{navy}is\_odd}|, A)

    @assert squares == [1, 4, 9, 16, 25, 36, 49, 64]
    @assert total_sum == 36
    @assert sum_squares == 204
    @assert odds == [1, 3, 5, 7]
end

Torc.|{\color{magenta}init}|(|{\color{navy}main}|)
\end{minted}
    \end{tcolorbox}
\vspace{-1.02em}
\captionof{listing}{Skeletal programming in Torcalia.}
\label{lst:skeletons}
\vspace{-1em}
\end{figure}

%% file: sections/evaluation.tex
\section{Performance Evaluation}
\label{sec:evaluation}

We evaluated the scalability of Torcalia across four task-parallel applications. Multiple benchmark runs were performed with different worker configurations (i.e., number of MPI ranks), while keeping
the number of Julia worker threads per rank fixed at 2; we report the median of five runs for each worker configuration. For the two workloads with irregular task granularity, the same experimental suite was also conducted using Dagger as the underlying runtime system, enabling direct performance comparison between the two tasking libraries. All code required to reproduce the experiments is publicly available at~\href{https://github.com/CEID-HPCLAB/Torcalia.jl}{\faGithub}.

\vspace{-0.3em}
\subsection{Experimental Setup}
\label{sec:setup}

\begin{figure*}[t!]
\centering
\includegraphics[width=1\textwidth]{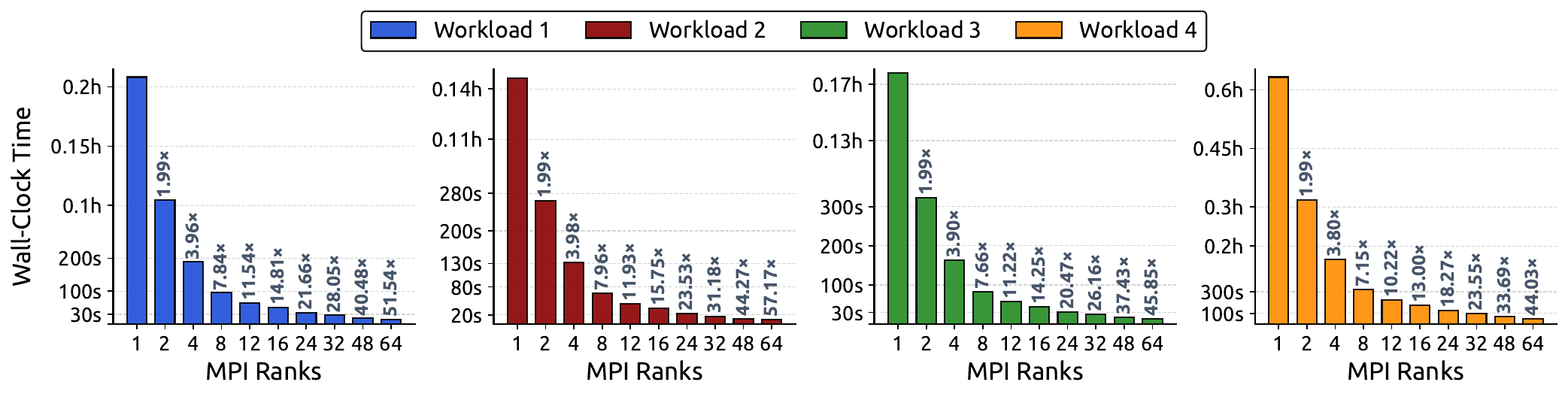}
\caption{Runtime performance of Torcalia for the four evaluated workloads. Across all workloads, increasing the number of MPI ranks from 1 to 64 results in a wall-clock time reduction of over 97\%.}
\vspace{-0.2cm}
\label{fig:bar_plot_times}
\end{figure*}

\vspace{0.2em}

\noindent\textbf{Workloads.} The four workloads span diverse application domains with varying task granularity. The \textbf{first} applies the Hooke--Jeeves algorithm~\cite{Hooke1961} to find the global minimum of the Rosenbrock benchmark function. Multiple independent runs, each starting from randomly selected initial points, are performed in parallel, enabling simultaneous exploration of different regions of the solution space. Specifically, the workload consists of 500 tasks, each executing 400 runs with a maximum of 5,000 iterations per run. The \textbf{second} uses Monte Carlo simulation to price a European call option. Stock price paths are simulated using Geometric Brownian Motion (GBM), and the option price is computed as the discounted average of the simulated payoffs. In total, $10^{11}$ iterations are performed, evenly distributed across 1,000 tasks. The \textbf{third} workload, commonly encountered in machine learning workflows, involves the parallel preprocessing of a large tabular dataset stored in CSV format (126~GB), following the task-based parallelism scheme proposed in~\cite{Mneme}. The input dataset is partitioned into 600 equally sized chunks, each processed by an individual task. Finally, the \textbf{fourth} application concerns the parallel Mandelbrot set computation, a well-known embarrassingly parallel benchmark with inherently irregular workload. The generated image has a fixed size of $5192\times5192$, with a maximum of 200,000 iterations per pixel, and each task processes a $64\times64$ pixel block, except at the image boundaries, where smaller blocks are formed.

\vspace{0.6em}

\noindent\textbf{Hardware Configuration.} All experiments were conducted on the ARIS supercomputer, a national HPC cluster facility, using four thin compute nodes. Each node is equipped with a dual-socket AMD EPYC 7763 CPU (128 cores, 2.45~GHz), partitioned into eight NUMA domains, with 512~GB of RAM. For all experiments, hyperthreading was disabled. The software stack includes Julia~v1.12.4, Dagger.jl~v0.22.7, and MPI.jl v0.20.29 with OpenMPI~v4.1.8.

\begin{figure}[b!]
\centering
\vspace{-0.5em}
\includegraphics[width=1\linewidth]{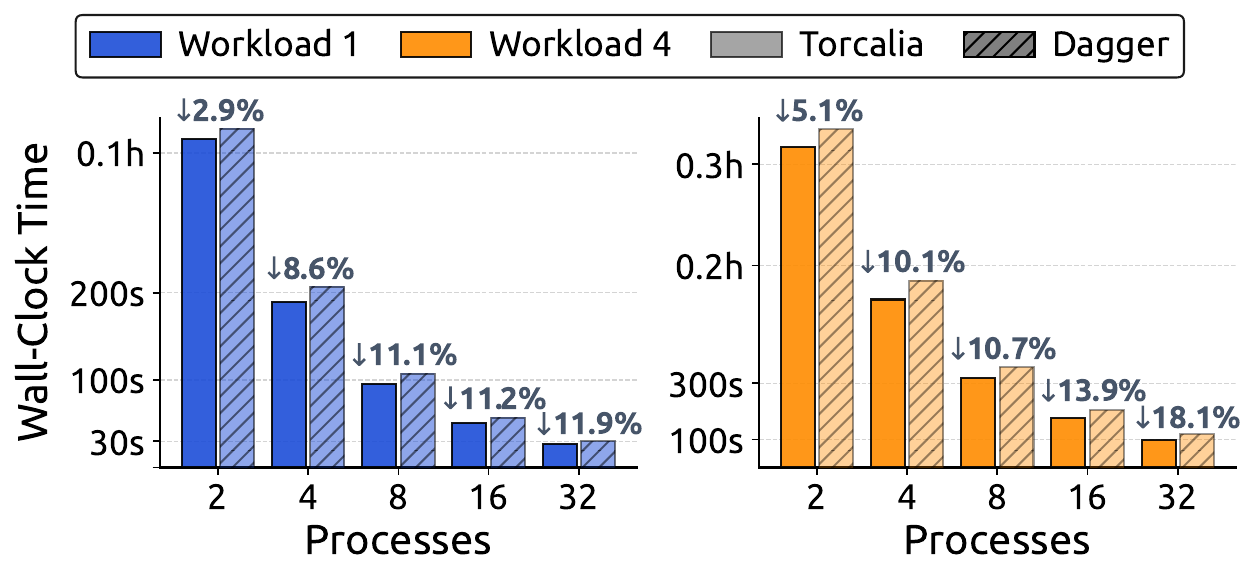}
\caption{Performance comparison with Dagger. The percentages above the bars indicate the reduction in wall-clock time achieved by Torcalia relative to Dagger for each worker configuration.}
\label{fig:bar_plot_dagger}
\vspace{-0.25em}
\end{figure}

\subsection{Results}
\label{sec:results}

Figure~\ref{fig:bar_plot_times} presents the runtime performance of Torcalia across the four workloads. The first and second workloads demonstrate strong scalability, with parallel efficiencies exceeding 80\% and 89\%, respectively, across all evaluated configurations. For the third workload, parallel efficiency remains above 85\% for configurations with fewer than 32 MPI processes and drops to 72\% at 64 processes. This degradation is primarily attributed to the high I/O demand associated with this workload, which exceeds the throughput that the disk subsystem can sustain as the number of workers increases. Regarding the fourth workload, increasing the number of MPI processes from 1 to 64 results in a wall-clock time reduction from 38 minutes to less than one minute, which corresponds to a 44.03$\times$ speedup. This speedup is the lowest among the four workloads, reflecting the irregular nature of the Mandelbrot set computation. Nevertheless, thanks to its \emph{work-stealing} mechanism, Torcalia adapts dynamically to the resulting load imbalance, keeping parallel efficiency above 85\% for configurations with fewer than 16 MPI processes and above 68\% for up to 64 MPI processes.

\vspace{0.55em}

\noindent\textbf{Comparison with Dagger.} As shown in Figure~\ref{fig:bar_plot_dagger}, Torcalia consistently outperforms Dagger across both workloads, reducing wall-clock time by 2.9\%--11.9\% for the first workload and 5.1\%--18.1\% for the fourth. Moreover, Torcalia demonstrates better scalability than Dagger. For the first workload, Torcalia maintains parallel efficiency above 87\% for up to 32 processes, whereas Dagger falls below 83\% already at 16 processes. A similar trend is observed for the fourth workload, where Torcalia retains parallel efficiency above 73\% up to 32 processes, whereas Dagger drops to 60.3\%.

\vspace{-0.3em}

%% file: sections/impact.tex
\section{Impact}
\label{sec:impact}

Designed to address the \textit{two-language problem}, Julia offers a powerful platform for developing high-performance scientific computing and machine learning applications. Such applications increasingly exhibit irregular, dynamically evolving workloads, for which the rigidity of conventional BSP models becomes a bottleneck~\cite{Joseph2026, cheng2026, Posner2025}, motivating the development of Julia-native AMT runtime systems~\cite{bantel2026, Dagger}, such as \textbf{Torcalia}. Built on top of MPI, Torcalia is a portable tasking library with adaptive load balancing, enabling researchers to scale their workflows from local workstations to multi-node clusters without modifying the application code. Moreover, its support for dynamic switching to the SPMD programming model allows existing MPI-based applications to incorporate task-based parallelism while preserving their established communication code, addressing a key barrier to adopting AMT runtimes in legacy HPC codebases~\cite{Posner2025}. Finally, by providing algorithmic skeletons, Torcalia further simplifies the implementation of parallel applications at scale~\cite{Herrmann2023}.

\vspace{0.6em}

\noindent\textbf{Target Audience.} Torcalia is primarily intended for two communities: (i)~computational scientists using Julia for large-scale simulations and other applications with irregular or adaptive workloads that would otherwise require complex, hand-written MPI code~\cite{Joseph2026}, and (ii)~developers of monolithic MPI applications seeking to incorporate dynamic load balancing incrementally, without completely rewriting or restructuring their communication layer~\cite{Posner2025}.

\vspace{0.6em}

\noindent\textbf{Application Domain.} The proposed tasking library is particularly useful for applications with recursive, irregular, nested, or adaptive parallelism, which are challenging to implement efficiently using static loop-based decompositions and BSP paradigms~\cite{HRPF,  Reitz2025}. Common examples include adaptive mesh refinement methods, graph-based computations, Markov Chain Monte Carlo (MCMC) algorithms, and computational pipelines in radio astronomy. For the latter, a complete example demonstrating parallel flux measurement of radio sources using Torcalia is available at~\href{https://github.com/CEID-HPCLAB/Torcalia.jl}{\faGithub}.

\vspace{0.6em}

\noindent\textbf{Research Impact.} As an open-source AMT runtime system, Torcalia can support the development of parallel domain-specific frameworks in Julia. For example, Mneme.jl~\cite{mnemeJulia} is built on top of Torcalia as a Julia counterpart to the Mneme Python package~\cite{Mneme}. Moreover, its low-level configuration mechanisms allow researchers to implement and evaluate new scheduling and elasticity policies. Finally, Torcalia can also be used for educational purposes, offering a flexible, unified environment for exploring various aspects of parallel computing, including scheduling policies, the impact of work stealing on irregular workloads, and the differences between the AMT and BSP programming models.

%% file: sections/conclusion.tex
\section{Conclusions}
\label{sec:conclusion}

We introduced Torcalia.jl, a Julia package for asynchronous task-based parallelism across shared and distributed memory platforms. At its core, the proposed runtime system adopts an MPI+X model, transparently leveraging MPI primitives for inter-node communication and Julia-native multithreading for intra-node parallelism. Task execution is orchestrated by a dependency-aware scheduler that combines \emph{work sharing} and \emph{work stealing} to provide adaptive load balancing. Tasks can be nested arbitrarily, supporting multi-level parallelism, while the integration of ArgoDSM enables task-based shared-memory programming on HPC clusters. Additional features include algorithmic skeletons, which simplify the development of parallel applications, and dynamic switching to the SPMD model, which allows task-based parallelism to be introduced incrementally into legacy MPI codebases.

Our ongoing work focuses on extending Torcalia to heterogeneous architectures, enabling tasks to be executed on GPUs and scheduled seamlessly alongside CPU tasks. Furthermore, we plan to introduce automatic elastic resource management, allowing computational resources to scale dynamically at runtime without user intervention. Additional directions include fault-tolerance mechanisms and refinements to the work-stealing strategy aimed at further optimizing task redistribution across MPI nodes.

%% file: sections/acks.tex
\section*{Acknowledgments}
\label{sec:acks}

This work was supported by computational time granted by the National Infrastructures for Research and Technology S.A. (GRNET S.A.) at the National HPC facility ARIS under project ID pa261001torcjulia.

%% file: sections/funding.tex
\section*{Funding}
\label{sec:funding}

This research was partially supported by project MIS 5154714, under the National Recovery and Resilience Plan Greece 2.0, funded by the European Union through the NextGenerationEU program, and by the easyHPC@eco.plastics.industry [6001593], which is co-funded by the European Union under the Competitiveness Program (ESPA 2021--2027). 